\documentclass{amsart}
\usepackage{amsfonts}
\usepackage{amsmath}
\usepackage{amssymb}
\usepackage{graphicx}

\begin{document}
\title[Reference Potential Approach to the Inverse Problem: I]{Reference potential approach to the quantum-mechanical inverse problem:
\linebreak I. Calculation of phase shift and Jost function}
\author{Matti Selg}
\address{Riia 142, 51014 Tartu, Estonia}
\email{matti@fi.tartu.ee}
\keywords{Inverse problem, Jost function, Levinson theorem}

\begin{abstract}
Elegant and mathematically rigorous methods of the quantum inverse theory are
difficult to put into practice because there is always some lack of needful
input information. In this situation, one may try to construct a reference
potential, whose spectral characteristics would be in a reasonable agreement
with the available data of the system's properties. Since the reference
potential is fixed, it is always possible to calculate all its spectral
characteristics, including phase shift for scattering states and Jost
function, the main key to solve the inverse problem. Thereafter, one can
calculate a Bargmann potential whose Jost function differs from the initial
one only by a rational factor. This way it is possible, at least in principle,
to construct a more reliable potential for the system. The model system
investigated in this paper is diatomic xenon molecule in ground electronic
state. Its reference potential is built up of several smoothly joined Morse
type components, which enables to solve the related energy eigenvalue problem
exactly. Moreover, the phase shift can also be calculated in part
analytically, and the Jost function can be acertained very accurately in the
whole range of positive energies. Full energy dependence of the phase shift
has been determined and its excellent agreement with the Levinson theorem
demonstrated. In addition, asymptotically exact analytic formulas for the
phase shift and the Jost function, independent of each other, are obtained and
their physical background elucidated.

\end{abstract}
\maketitle

\section{Introduction}

Strict mathematical criteria for the unique solution of the inverse problem
for the one-dimensional Schr\"{o}dinger equation have been formulated more
than 50 years ago. The history of this prestigious research area goes back to
the dawn of quantum mechanics \cite{Ambartsumyan}, and the decisive
breakthrough has been achieved thanks to the important contributions by Borg
\cite{Borg, Borg52}, Levinson \cite{Levinson1, Levinson2}, Bargmann
\cite{Bargmann1, Bargmann2}, Gel'fand and Levitan \cite{GL}, Jost and Kohn
\cite{Jost1, Jost2}, Marchenko \cite{Marchenko1, Marchenko2}, Krein
\cite{Krein1, Krein2}, and others (see, e.g., \cite{Chadan} for a thorough
overview). As a result of fruitful brainstorming which culminated in early
fifties it has been established that the quantum mechanical inverse problem
can be solved if one manages to fix the so-called spectral function
\cite{Marchenko1}. In the case of confining one-dimensional quantum systems it
means that the interaction potential can be uniquely determined if and only if
the following complete set of information is available:

\begin{itemize}
\item Full energy spectrum of the bound states $E_{n}$ $(n=0,1,...,N).$

\item Full energy dependence (from 0 to $\infty$) of the phase shift
$\delta(E)$ for the scattering states $E>0$.

\item $n$ additional real parameters $C_{n}$ $(n=0,1,...,N)$ related to
relevant bound states that uniquely fix their normalization.
\end{itemize}

In principle, all these data can be obtained experimentally, but
unfortunately, this is almost unachievable in practice. The real situation is
even more hopeless, because in addition to the deficit of input information
one inevitably faces very serious computational-technical problems. Thus, in
spite of the whole mathematical beauty of the theory, one comes to a
regrettable conclusion that rigorous solution of the quantum-mechanical
inverse problem is a tremendously difficult task. For this reason most methods
of deducing potentials from the available experimental data are based on some
simplifying preconditions, therefore being inaccurate from the rigorous
quantum mechanical point of view. Nevertheless, such methods may prove quite
useful. For example, semiclassical approaches introduced many years ago by
Rydberg \cite{Ryd1, Ryd2}, Klein \cite{Klein}, Rees \cite{Rees} and Dunham
\cite{Dunham1, Dunham2}\ are still very popular in spectroscopy of diatomic
molecules, and these concepts are constantly improved and developed
\cite{LeRoy1, LeRoy2}.

Is it possible to apply rigorous methods of the inverse quantum theory for
practical purposes? In view of the principle difficulties mentioned above, one
has to be cautious in answering this question. In this paper we restrict
ourselves to simple one-dimensional quantum systems, and the analysis proceeds
from an idea that for any system of this kind one can build up a reasonable
reference potential based on the available experimental data. As the reference
potential is known, it is always possible to calculate all its discrete energy
eigenvalues $E_{n}$, their norming constants $C_{n}$, and the phase shifts
$\delta(E)$ for the scattering states. Therefore, in this artificial way one
gets the full set of input information needed to uniquely solve the quantum
mechanical inverse problem. Of course, the described approach is tautological:
there is no need to regain a potential which is already known by definition.
Nevertheless, such an approach is not meaningless, as it gives good zeroth
approximations to the important spectral characteristics, such as \textit{Jost
function} and \textit{spectral density} (the terms to be specified below). One
can interpret the reference potential as only an initial guess to the real
potential. Although the calculated quantities $E_{n}$, $C_{n}$ and $\delta(E)$
do not exactly match the actual values for the real system, they are still
expected to be quite close to them. For a given reference potential one can
calculate its Bargmann potential whose Jost function differs from the initial
one only by a rational factor \cite{Chadan}. By a suitable choice of this
factor one can take a more adequate account of the experimental data. For
example, one can replace the calculated discrete energy eigenvalues $E_{n}$
related to the reference potential with their actually observed values.
Consequently, at least in some sense the new potential would be more realistic
than the initial reference potential. There is also another motivation for the
described "inverse" approach to the inverse problem. Namely, through direct
practical experience one can essentially increase his knowledge of how to
overcome serious computational-technical difficulties when applying rigorous
methods of the inverse quantum theory.

The one-dimensional inverse theory can be applied to diatomic molecules, since
a two-particle problem can be always reduced to a one-particle problem in a
spherically symmetric field. On this basis, we are going to examine the
inverse problem for diatomic xenon molecule in its ground electronic state.
Several reports of the research are planned. Methods of solution of the
integral equations that enable to uniquely ascertain the potential are
discussed and illustrated in the next paper of this series \cite{SelgII},
while in this paper the emphasis is put on explaining the details of the basic
concepts related to the reference potential approach. In Section 2 we briefly
describe how the reference potential for the model system has been
constructed, and how its discrete energy eigenvalues have been calculated. In
Section 3 we describe the details of calculating the phase shift for the
scattering states, and demonstrate full agreement with Levinson theorem
\cite{Levinson1}. Section 4 aims to explain the important role of the Jost
function in the quantum inverse theory. In particular, a detailed analysis of
the asymptotic behavior of the Jost function is given. Finally, a brief
conclusion is given in Section 5.

\section{Exactly solvable reference potential for Xe$_{2}$}

In Fig. 1 one can see the reference potential constructed for the Xe$_{2}$
molecule. The same curve is depicted in both graphs, but very different energy
scales are used. Throughout this paper only the rotationless case is analyzed,
i.e., the rotational quantum number $J=0.$ According to the starting idea of
the approach the only criterion for the choice of the reference potential is
its agreement with the available experimental data. Therefore, we will not pay
too much attention to various mathematical nuances and simply assume that the
reference potential should be smooth and integrable in the whole physical
domain. In addition, we try to construct a reference potential whose analytic
form is as simple as possible. A good choice for this purpose, as explained in
detail elsewhere \cite{Selg1, Selg2}, is a multi-component potential composed
of smoothly joined Morse-type pieces%
\begin{equation}
V(r)=V_{k}+D_{k}\left[  \exp(-\alpha_{k}(r-r_{k}))-1\right]  ^{2},\text{ }%
r\in(0,\infty),
\end{equation}
where $V_{k},$ $D_{k},$ $\alpha_{k}$ and $r_{k}$ are some constants (not
definitely positive), and the subscript $k$ corresponds to different
components smoothly joined at some suitably chosen boundary points $X_{k+1}$.
The reference potential shown in Fig. 1 consists of only three components
($k=0,1,2$), the most internal of them ($k=0$) being a so-called pseudo-Morse
potential. It means that the tiny potential well corresponding to this
component (if taken separately) is just of the limit depth to entirely lose
the discrete energy spectrum. Consequently, $D_{0}=\hbar^{2}\alpha_{0}%
^{2}/(8m)$ ($m$ being the reduced mass of the pair of atoms), so that only
three independent parameters remain for this component. The central component
($k=1$) is an ordinary Morse potential, while the most external one ($k=2$) is
a "reversed" Morse potential with the parameter $D_{2}$ being negative. By
introducing a "reversed" component one artificially creates a small potential
hump in the long-distance range. This might seem unphysical and unjustified,
but the point is that the height of this artificial hump approaches zero as
the parameter $r_{2}$\ approaches infinity. Therefore, taking a sufficiently
large $r_{2}$, the hump becomes almost insignificant, while the analytic
treatment remains simple and flexible. All parameters of the reference
potential can be easily determined, if one requires continuity of the
potential and its first derivative at the boundary points $X_{1}$ and $X_{2}$
(also shown in Fig. 1). These parameters as well as the calculated discrete
energy eigenvalues $E_{n}$ $(n=0,1,...,23)$ are given in Table 1.

The essence of the described construction is that the energy eigenvalue
problem for the reference potential can be easily solved by solely analytic
means to any desired accuracy \cite{Selg1}. Moreover, as we demonstrate in the
next Section, the major part of the energy dependence of the phase shift can
also be ascertained analytically, which is a great advantage compared with
applying numerical methods.

Most spectroscopic applications are related to the distance region shown in
the lower graph of Fig. 1. For the inverse quantum theory, however, it is
important to accurately reproduce the potential near the zero point $r=0,$
which is almost meaningless for spectroscopic applications. In this context
one cannot ignore the fact that according to Eq. (1) the reference potential
is finite at $r=0$ (see the upper graph of Fig. 1). Actually it means that the
potential "jumps" to infinity at zero point, and this might also seem
unphysical and unjustified. However, one has to bear in mind that the behavior
of the real potential near $r=0$ is unknown and remains unknown. It does not
matter so much how we describe the potential in this region, in so far as
spectroscopic applications are of our main interest. As we see in the next
section, description in terms of a pseudo-Morse potential is mathematically
simple and elegant, and this, too, can be taken as a motivation for the
approach to be used.

\section{Phase shift and Levinson theorem}

Next we are going to calculate the phase shift $\delta(E)$ for the full range
of scattering states $E\in(0,\infty).$ To this end, one can use a long-known
method first introduced by Morse and Allis \cite{Morse}(let us remind that
only the rotationless case $J=0$ is examined here). It is based on solution of
the following equation:%
\begin{equation}
\delta^{\prime}(r,k)=-\dfrac{\sqrt{2m}V(r)}{\hbar k}\sin^{2}\left[
kr+\delta(r,k)\right]  ,\text{ }\delta(0,k)=0,
\end{equation}
where $k=$ $\dfrac{\sqrt{2mE}}{\hbar}.$ The phase shift is then determined as
$\delta(k)=\lim\limits_{r\rightarrow\infty}\delta(r,k).$ The described method
is universal but rather time-consuming, because for any energy from 0 to
$\infty$ one has to perform an integration from 0 to $\infty.$ Fortunately, at
this point we can take advantage of the special analytic form of the reference
potential. As mentioned, the region $r\in(0,X_{1})$ is approximated by a
pseudo-Morse potential. It means that one immediately gets two linearly
independent analytic solutions of the corresponding Schr\"{o}dinger equation
\cite{Selg1, Selg2}%
\begin{align}
\Psi_{0}^{(1)}(r)  &  =\exp(-y_{0}/2)y_{0}^{i\mu_{0}}\Psi(i\mu_{0},2i\mu
_{0}+1;y_{0})\\
\Psi_{0}^{(2)}(r)  &  =\exp(y_{0}/2)(-y_{0})^{i\mu_{0}}\Psi(i\mu_{0},2i\mu
_{0}+1;-y_{0}),\text{ }r\in(0,X_{1}),\nonumber
\end{align}
where $\mu_{0}=1/2\sqrt{(E-V_{0})/D_{0}-1}>0$, $y_{0}\equiv\exp\left[
-\alpha_{0}(r-r_{0})\right]  $, and $\Psi(i\mu_{0},2i\mu_{0}+1;y_{0})$ is a
particular solution of the confluent hypergeometric equation introduced by
Tricomi (see \cite{Bateman}\ for details). If $\mu_{0}^{2}<<y_{0},$ this
function can be evaluated from the asymptotic series
\begin{equation}
\Psi(i\mu_{0},2i\mu_{0}+1;y_{0})=y_{0}^{-i\mu_{0}}\sum_{n=0}^{N}\frac
{(i\mu_{0})_{n}(-i\mu_{0})_{n}}{n!(-y_{0})^{n}},
\end{equation}
where $(a)_{n}\equiv\Gamma(a+n)/\Gamma(a)=a(a+1)(a+2)...(a+n-1)$ is the
Pochhammer symbol, and $N$ must not be too large. Thus%
\begin{equation}
\Psi_{0}^{(1)}(r)=\exp(-y_{0}/2)\left[  1-\frac{\mu_{0}^{2}}{1!y_{0}}%
+\frac{\mu_{0}^{2}(\mu_{0}^{2}+1^{2})}{2!y_{0}^{2}}-\frac{\mu_{0}^{2}(\mu
_{0}^{2}+1^{2})(\mu_{0}^{2}+2^{2})}{3!y_{0}^{3}}+...\right]  ,
\end{equation}
and analogously%
\begin{equation}
\Psi_{0}^{(2)}(r)=\exp(y_{0}/2)\left[  1+\frac{\mu_{0}^{2}}{1!y_{0}}+\frac
{\mu_{0}^{2}(\mu_{0}^{2}+1^{2})}{2!y_{0}^{2}}+\frac{\mu_{0}^{2}(\mu_{0}%
^{2}+1^{2})(\mu_{0}^{2}+2^{2})}{3!y_{0}^{3}}+...\right]  .
\end{equation}

The phase shift is related to \bigskip\textit{regular solutions} of the
Schr\"{o}dinger equation, which means that the physically correct linear
combination of$\ \Psi_{0}^{(1)}(r)$ and $\Psi_{0}^{(2)}(r)$ should vanish as
$r\rightarrow0$, i.e.,%
\[
\Psi_{0}(r)=N_{1}\Psi_{0}^{(1)}(r)+N_{2}\Psi_{0}^{(2)}(r),
\]
where%
\begin{equation}
\dfrac{N_{2}}{N_{1}}=-\exp\left[  -y_{0}(0)\right]  \cdot\dfrac{1-\dfrac
{\mu_{0}^{2}}{1!y_{0}(0)}+\dfrac{\mu_{0}^{2}(\mu_{0}^{2}+1^{2})}{2!y_{0}%
^{2}(0)}-...}{1+\dfrac{\mu_{0}^{2}}{1!y_{0}(0)}+\dfrac{\mu_{0}^{2}(\mu_{0}%
^{2}+1^{2})}{2!y_{0}^{2}(0)}+...}.
\end{equation}

From Eq. (7) one can see that the particular solution $\Psi_{0}^{(2)}(r)$
practically does not contribute to the regular solution at distances
sufficiently far from zero point, but still deep inside the classically
forbidden region. Indeed, in this case%
\begin{gather}
\dfrac{N_{2}\Psi_{0}^{(2)}(r)}{N_{1}\Psi_{0}^{(1)}(r)}=-\exp\left[
y_{0}(r)-y_{0}(0)\right]  \times\dfrac{1-\dfrac{\mu_{0}^{2}}{1!y_{0}%
(0)}+\dfrac{\mu_{0}^{2}(\mu_{0}^{2}+1^{2})}{2!y_{0}^{2}(0)}-...}{1+\dfrac
{\mu_{0}^{2}}{1!y_{0}(0)}+\dfrac{\mu_{0}^{2}(\mu_{0}^{2}+1^{2})}{2!y_{0}%
^{2}(0)}+...}\times\\
\times\dfrac{1+\dfrac{\mu_{0}^{2}}{1!y_{0}(r)}+\dfrac{\mu_{0}^{2}(\mu_{0}%
^{2}+1^{2})}{2!y_{0}^{2}(r)}+...}{1-\dfrac{\mu_{0}^{2}}{1!y_{0}(r)}+\dfrac
{\mu_{0}^{2}(\mu_{0}^{2}+1^{2})}{2!y_{0}^{2}(r)}-...}=-\exp\left[
y_{0}(r)-y_{0}(0)\right]  \times\nonumber\\
\times\dfrac{1+\dfrac{\mu_{0}^{2}}{1!y_{0}(0)}\left[  \exp(\alpha
_{0}r)-1\right]  +...}{1-\dfrac{\mu_{0}^{2}}{1!y_{0}(0)}\left[  \exp
(\alpha_{0}r)-1\right]  +...}\approx-\exp\left\{  -\exp(\alpha_{0}%
r_{0})\left[  1-\exp(-\alpha_{0}r)\right]  \right\} \nonumber
\end{gather}
is an extremely small quantity.

There is still a lot more profit to gain from the pseudo-Morse approximation
to calculate the phase shift. As we just proved, the physically correct
solution (apart from normalization) in a wide energy range (practically up to
$E\approx V(0)$) reduces to the particular solution $\Psi_{0}^{(1)}(r).$ Using
some well-known relations from the theory of confluent hypergeometric
functions \cite{Bateman, Selg1}\, this solution can be rewritten%
\begin{gather}
\Psi_{0}^{(1)}(r)=\exp(-y_{0}/2)y_{0}^{i\mu_{0}}\Psi(i\mu_{0},2i\mu
_{0}+1;y_{0})=\\
\exp(-y_{0}/2)\left[  \frac{\Gamma(-2i\mu_{0})}{\Gamma(-i\mu_{0})}y_{0}%
^{i\mu_{0}}\Phi(i\mu_{0},2i\mu_{0}+1;y_{0})+\frac{\Gamma(2i\mu_{0})}%
{\Gamma(i\mu_{0})}y_{0}^{-i\mu_{0}}\Phi(-i\mu_{0},-2i\mu_{0}+1;y_{0})\right]
,\nonumber
\end{gather}
where the symbols $\Phi(a,c;x)=1+\dfrac{a}{c}\dfrac{x}{1!}+\dfrac
{a(a+1)}{c(c+1)}\dfrac{x^{2}}{2!}+...$ denote the well-known confluent
hypergeometric functions.

As is seen, Eq. (9) represents a sum of two complex conjugates. Therefore,%
\[
\Psi_{0}^{(1)}(r)\sim A_{0}(y_{0})\cos\left[  B_{0}(y_{0})-\varphi_{0}%
-\alpha_{0}\mu_{0}r\right]  ,
\]
where the functions $A_{0}$ and $B_{0}$ can be determined from a series
\cite{Selg1}:
\begin{align}
A_{0}(y_{0})e^{iB_{0}(y_{0})}  &  =1-\frac{y_{0}/4}{i\mu_{0}+1/2}%
+\frac{\left(  y_{0}/4\right)  ^{2}}{\left(  i\mu_{0}+1/2\right)  1!}\left(
1-\frac{y_{0}/4}{i\mu_{0}+3/2}\right)  +\nonumber\\
+  &  \frac{\left(  y_{0}/4\right)  ^{4}}{\left(  i\mu_{0}+1/2\right)  \left(
i\mu_{0}+3/2\right)  2!}\left(  1-\frac{y_{0}/4}{i\mu_{0}+5/2}\right)  +...
\end{align}
A good point is that the phase parameter (which, of course, is not the actual
phase shift) $\varphi_{0}\equiv$ $\alpha_{0}\mu_{0}r_{0}-\arg\left[
\dfrac{\Gamma(2i\mu_{0})}{\Gamma(i\mu_{0})}\right]  $ can be calculated
exactly. Indeed, using a Legendre formula for doubling a gamma function's
argument \cite{Bateman}%
\[
\Gamma(2z)=\frac{2^{2z-1}}{\sqrt{\pi}}\Gamma(z)\Gamma(z+1/2),
\]
and another useful formula \cite{Cheb}%
\begin{equation}
\arg\Gamma(i\mu_{0}+1/2)=\mu_{0}\left(  \frac{1}{2}\ln(1+4\mu_{0}^{2}%
)-1-\ln2\right)  -\frac{1}{2}\int\limits_{0}^{\infty}\left(  \coth t-\frac
{1}{t}\right)  e^{-t}\sin(2\mu_{0}t)\frac{dt}{t},
\end{equation}
one gets the following result:%

\begin{equation}
\varphi_{0}=\mu_{0}\left(  \alpha_{0}r_{0}+1-\ln2-\frac{1}{2}\ln(1+4\mu
_{0}^{2})\right)  +\frac{1}{2}\int\limits_{0}^{\infty}\left(  \coth t-\frac
{1}{t}\right)  e^{-t}\sin(2\mu_{0}t)\frac{dt}{t},
\end{equation}
where the integral can be conveniently evaluated \cite{Selg1}%
\begin{equation}
I\equiv\int\limits_{0}^{\infty}\left(  \coth t-\frac{1}{t}\right)  e^{-t}%
\sin(2\mu_{0}t)\frac{dt}{t}=\int\limits_{0}^{T}e^{-t}\sin(\pi\frac{t}%
{T})f(t)dt,\text{ }T=\frac{\pi}{2\mu_{0}},
\end{equation}%
\[
f(t)=\frac{\coth t-\frac{1}{t}}{t}-e^{-T}\frac{\coth(t+T)-\frac{1}{t+T}}%
{t+T}+e^{-2T}\frac{\coth(t+2T)-\frac{1}{t+2T}}{t+2T}-...
\]
Another equivalent formula for this quantity reads \cite{Selg2}%

\begin{equation}
I=\sum_{n=1}^{\infty}I_{n},\text{ \ }I_{n}=\frac{(-1)^{n-1}2^{2n}B_{n}%
}{(2n)(2n-1)(1+4\mu_{0}^{2})^{2n-1}}\sum_{k=0}^{n-1}(-1)^{k}\binom{2n-1}%
{2k+1}(2\mu_{0})^{2k+1}.
\end{equation}
Here $B_{n}$ denotes the $n$-th order Bernoulli number.

Correct linear combinations of solutions of the Schr\"{o}dinger equations for
analytically different pieces of the reference potential are uniquely fixed by
the continuity requirements of the wave function and its first derivative at
the boundary points $X_{1}$ and $X_{2}$. Since the parameter $\varphi_{0}$ can
be calculated exactly, the phase shift can also be ascertained with the help
of solely analytic means in the energy range $0<E\lesssim V(0).$ At high
energies ($E\gtrsim V(0)$), however, the analytic approach gradually fails,
since the particular solution $\Psi_{0}^{(2)}(r)$\ in Eq. (3) cannot be
ignored any more. To determine the phase shift in this region one has to solve
Eq. (2). Fortunately, there is no need to perform integration over the whole
physical domain $r\in(0,\infty)$, because the solution in the long-distance
range $r\geq X_{2}$ already has the "right" analytic form \cite{Selg1}%
\begin{equation}
\Psi_{2}(r)=2A_{2}(y_{2})\cos\left[  B_{2}(y_{2})+\varphi_{2}-kr\right]
=2A_{2}(y_{2})\sin\left[  \delta(r,k)+kr\right]  ,
\end{equation}
from which the phase shift can be easily obtained. Here $y_{2}(r)\equiv
2a_{2}\exp\left[  -\alpha_{2}(r-r_{2})\right]  ,$ $a_{2}\equiv\sqrt{2mD_{2}%
}/\left(  \hbar\alpha_{2}\right)  ,$ $\delta(r,k)$ is the solution of Eq.
(2), and the complex function $A_{2}(y_{2})\exp\left[  iB_{2}(y_{2})\right]
\equiv\exp\left(  -iy_{2}/2\right)  \Phi\left[  i(k/\alpha_{2}-a_{2}%
)+1/2,2ik/\alpha_{2}+1;iy_{2}\right]  .$ Since $A_{2}(y_{2})\rightarrow0$ and
$B_{2}(y_{2})\rightarrow1$ as $r\rightarrow0,$ the phase shift%
\begin{equation}
\delta(k)=\delta(X_{2},k)+B_{2}\left[  y_{2}(X_{2})\right]  ,
\end{equation}
i.e., one only needs to integrate Eq. (2) until the boundary point $X_{2}$.

Now, let us recall an important relation known as Levinson theorem
\cite{Levinson1}%
\begin{equation}
\delta(0)-\delta(\infty)=n\pi,
\end{equation}
which correlates the energy dependence of the phase shift with the number of
bound states. As can be seen in Fig. 2, a really good agreement with the
Levinson theorem can be obtained, but only if the phase shift is calculated up
to very high energies (note that the energy scale in Fig. 2 is logarithmic,
and it involves 20 orders of magnitude!). Moreover, the phase shift has to be
calculated with sufficiently high precision throughout the whole energy range,
otherwise there is no chance to accurately ascertain other important spectral
characteristics, such as Jost function. Unfortunately, the higher the energy
goes, the more complicated and time-consuming the numerical integration of Eq.
(2) becomes. How could we bridge over this troublesome technical difficulty?
In such situation, one may recall some general principles, and this is indeed
helpful to complete calculations of the phase shift.

Let us express the phase shift as formal result of integration of Eq. (2):%
\begin{gather}
\delta(k)=\lim_{r\rightarrow\infty}\delta(r,k)=-\frac{1}{2Ck}\int
\limits_{0}^{\infty}V(r)\left\{  1-\cos\left[  2kr+2\delta(r,k)\right]
\right\}  dr=-\frac{\int\limits_{0}^{\infty}V(r)dr}{2Ck}+\\
+\frac{1}{2Ck}\int\limits_{0}^{\infty}\cos\left(  2kr\right)  \left\{
V(r)\cos\left[  2\delta(r,k)\right]  \right\}  dr-\frac{1}{2Ck}\int
\limits_{0}^{\infty}\sin\left(  2kr\right)  \left\{  V(r)\sin\left[
2\delta(r,k)\right]  \right\}  dr.\nonumber
\end{gather}
Here a special denotation has been introduced (and will be used henceforward)
for the constant $C\equiv\dfrac{\hbar^{2}}{2m}$ that often appears in
formulas. Now, let us call for help from the famous Riemann-Lebegue theorem
(see, e.g., \cite{Titchmarsh}): if a function $F(r)$ is integrable in an
interval $r\in(a,b)$, then
\begin{equation}
\int\limits_{a}^{b}\cos\left(  \lambda r\right)  F(r)dr\rightarrow0,\text{
}\int\limits_{a}^{b}\sin\left(  \lambda r\right)  F(r)dr\rightarrow0\text{ as
}\lambda\rightarrow\infty.
\end{equation}
From this one immediately concludes that the last two integrals in Eq. (18)
will vanish as $k\rightarrow\infty.$ However, we can integrate by parts and
apply the Riemann-Lebegue theorem to the resulting integrals. This procedure
can be repeated as many times as needed, and as a result, one comes to a
rather general asymptotic formula for the phase shift (it can easily proved
that only odd powers of $k$ appear in this series)%
\begin{equation}
\delta(k)=\dfrac{a_{1}}{k}+\frac{a_{3}}{k^{3}}+\frac{a_{5}}{k^{5}}+...,\text{
}k\rightarrow\infty,
\end{equation}
where%
\begin{equation}
a_{1}=-\frac{\int\limits_{0}^{\infty}V(r)dr}{2Ck},\text{ }a_{3}=-\frac
{1}{8C^{2}}\left\{  \int\limits_{0}^{\infty}\left[  V(r)\right]
^{2}dr+CV^{\prime}(0)\right\}  ,\text{ }%
\end{equation}%
\[
a_{5}=-\frac{1}{32C^{3}}\left\{  \int\limits_{0}^{\infty}\left[  V(r)\right]
^{3}dr+C\int\limits_{0}^{\infty}\left[  V^{\prime}(r)\right]  ^{2}%
dr+4CV(0)V^{\prime}(0)-C^{2}V^{\prime\prime\prime}(0)\right\}  .
\]
Thus one can easily calculate the coefficients $a_{1}$, $a_{3}$, $a_{5}$,
etc., and this is just what is needed to ascertain the whole energy dependence
of the phase shift. Fig. 3 demonstrates how well Eq. (20) fits with the
results of direct numerical integration of the phase equation.

One cannot so easily find any direct illustrations to the Levinson theorem
from the literature (at least the author of this paper has not found them),
and Fig. 2, which is an illustration of this kind, could therefore be of more
general interest than merely an attachment to a particular model potential.
The curve shown in this figure has been calculated with at least 8 significant
digits, and it is in full agreement with all relevant general physical
considerations. This concerns not only the Levinson theorem and the asymptotic
behavior at large energies, but also the low energy part of the energy
dependence (see the left-side inset of Fig. 2), exactly corresponding to the
well-known formula%
\begin{equation}
\delta(k)=n\pi-\arctan(kR_{0}),\text{ }k\rightarrow0
\end{equation}
($R_{0}$ being scattering length), which can be found in most handbooks on
quantum mechanics.

\section{Jost function and inverse problem}

Having calculated the phase shift, we have come to a situation from which an
ideal inverse problem study would start. In other words, we are now provided
with the full set of information needed to uniquely solve the inverse problem.
In our case this would mean that we simply regain the reference potential from
which we started. This, of course, is not our main goal. As explained in
Section 1, we are interested in getting realistic zeroth approximations to the
important spectral functions which then could be used to improve the initial
model potential. The latter step, however, is planned as a subject for a
forthcoming publication. In this paper we only try to make sure that the
described scheme is reliable, and to this end the next step is to accurately
calculate the Jost function, the main spectral characteristic, which contains
the most part of information needed to solve the inverse problem. A thorough
overview of all useful properties of the Jost function is given in Ref.
\cite{Chadan}. For the treatment here, the most important point is that the
Jost function creates a simple link between \textit{physical} and regular
solutions of the scattering states. The physical solution ($J=0$) reads (cf.
with Eq. (15))%
\begin{equation}
\Psi(r,k)\approx\exp i\delta(k)\sin\left[  kr+\delta(k)\right]  ,\text{
}r\rightarrow\infty,
\end{equation}
while the regular solution $\varphi(r,k)$ is defined by a condition%
\begin{equation}
\lim_{r\rightarrow0}\frac{\varphi(r,k)}{r}=1.
\end{equation}
It can be shown that these two solutions of the Schr\"{o}dinger equation are
proportional%
\begin{equation}
\Psi(r,k)=\frac{k}{F(k)}\varphi(r,k),
\end{equation}
where
\begin{equation}
F(k)=\left\vert F(k)\right\vert \exp\left[  -i\delta(k)\right]
\end{equation}
is the Jost function we are talking about. For further treatment we have to
calculate the modulus of this function \cite{Chadan}%
\begin{equation}
\left\vert F(E)\right\vert =\prod\limits_{n=0}^{N}(1-E_{n}/E)\exp\left[
-\frac{1}{\pi}P\int\limits_{0}^{\infty}\frac{\delta(E^{\prime})dE}{E^{\prime
}-E}\right]  ,\text{ }E\in(0,\infty).
\end{equation}
Here $E_{n}$ are the discrete energy eigenvalues and the symbol $P$ points at
the principal value of the integral. As the phase shift and the bound states
are known, the calculations are relatively trivial, but they have to be
carried out very accurately. Next we can fix the spectral density%
\begin{equation}
\dfrac{d\rho(E)}{dE}=\left\{\genfrac{}{}{0pt}{}{\pi^{-1}\sqrt{E}\left\vert F(E)\right\vert 
^{-2},\text{{}}E\geq0,}{\sum\limits_{n}C_{n}\delta(E-E_{n}),\text{ \ }E<0.}\right.
\end{equation}
Here $C_{n}$ are the norming constants for the relevant bound states. Note
that these quantities are related to regular solutions, wherefore their
ascertainment is not so easy task as one might think.

Now we have come close to the real solution schemes of the inverse problem.
For example, the Gelfand-Levitan method \cite{GL} is based on the integral
equation%
\begin{equation}
K(r,r^{\prime})+G(r,r^{\prime})+\int\limits_{0}^{r}K(r,s)G(s,r^{\prime})ds=0,
\end{equation}
whose kernel reads%
\begin{equation}
G(r,r^{\prime})=\int\limits_{-\infty}^{\infty}\frac{\sin\left(  kr\right)
\cdot\sin\left(  kr^{\prime}\right)  }{k^{2}}d\sigma,
\end{equation}
and the quantity $d\sigma\equiv d\rho(E)-\dfrac{d\rho_{0}(E)}{dE}dE.$ Here
$\dfrac{d\rho_{0}(E)}{dE}$ is free particle's spectral density (related to the
potential $V(r)\equiv0$), and therefore,%

\begin{equation}
d\sigma=\left\{\genfrac{}{}{0pt}{}{d\rho(E)-d(\dfrac{2E^{3/2}}{3\pi}),\text{ }E\geq
0}{d\rho(E),\text{ \ \ \ \ \ \ \ \ \ \ \ \ \ \ \ \ }E<0.}\right.
\end{equation}
Eq. (30) can be rewritten \cite{Chadan}%
\begin{equation}
G(r,r^{\prime})=\frac{2}{\pi}\int\limits_{0}^{\infty}\sin\left(  kr\right)
\cdot\sin\left(  kr^{\prime}\right)  g(k)dk+\sum_{n}\frac{C_{n}}{4\gamma
_{n}^{2}}\sinh\left(  \gamma_{n}r\right)  \sinh\left(  \gamma_{n}r^{\prime
}\right)  ,
\end{equation}
where $\gamma_{n}^{2}=-\dfrac{2mE_{n}}{\hbar^{2}}$ and the function%
\begin{equation}
g(k)\equiv\dfrac{1}{\left\vert F(k)\right\vert ^{2}}-1.
\end{equation}
If one is able to solve Eq. (29), the potential can be determined from the
relation%
\begin{equation}
V(r)=2C\frac{d}{dr}K(r,r).\text{ }%
\end{equation}

Eqs. (29) to (34) explicitly demonstrate great importance of the Jost
function in the inverse quantum theory. The energy dependence of the Jost
function's modulus is shown in Fig. 4. One can see that at small energies this
quantity achieves extremely large values. Near the "critical" energy $E=V(0)$
the curve rapidly turns from nearly vertical to nearly horizontal, and at
still higher energies it slowly approaches the limit value $\left\vert
F(k)\right\vert \rightarrow1$, "breaking free" from the potential field (note
that for a free particle, $F(k)=1$ independent of energy).

The kernel of the Gelfand-Levitan equation is essentially determined by the
function $g(k)$ given by Eq. (33), which means that this function has to be
ascertained very accurately, in order to solve Eq. (29) and calculate the
potential according to Eq. (34). As can be seen in Fig. 5, $g(k)$ noticeably
differs from unity only at $k\gtrsim k_{0}\equiv\sqrt{V(0)/C}$, and
$g(k)\rightarrow0$ as $k\rightarrow\infty.$ Naturally, as shown in the inset,
there is no break of derivative in the "critical" region. Since the asymptotic
behavior of the function $g(k)$ essentially determines the shape of the
potential near the zero point $r=0$ (see the end of this section), it makes
sense to analyze this behavior in more details. We have already ascertained
the asymptotic expression for the phase shift (see Eq. (20)), and this can
be used to immediately get the asymptotic formulas for both the Jost
function's modulus and the function $g(k)$. For example, taking%
\begin{equation}
\ln\left\vert F(k)\right\vert =\dfrac{a_{2}}{k^{2}}+\frac{a_{4}}{k^{4}}%
+\frac{a_{6}}{k^{6}}+...,\text{ }k\rightarrow\infty
\end{equation}
(this time only even powers of $k$ appear in the series, as can be easily
proved), the coefficients read%
\begin{align}
a_{2} &  =-\frac{2}{\pi}(a_{1}k_{a}-\frac{a_{3}}{k_{a}}-\frac{a_{5}}%
{3k_{a}^{3}}-...)+\frac{1}{C}\left[  \frac{1}{\pi}\int\limits_{0}^{E_{a}%
}\delta(E^{\prime})dE^{\prime}-\sum_{n}E_{n}\right]  ,\\
a_{4} &  =-\frac{2}{\pi}(\frac{a_{1}k_{a}^{3}}{3}+a_{3}k_{a}-\frac{a_{5}%
}{k_{a}}-...)+\frac{1}{C^{2}}\left[  \frac{1}{\pi}\int\limits_{0}^{E_{a}%
}\delta(E^{\prime})E^{\prime}dE^{\prime}-\frac{1}{2}\sum_{n}\left(
E_{n}\right)  ^{2}\right]  ,\nonumber\\
a_{6} &  =-\frac{2}{\pi}(\frac{a_{1}k_{a}^{5}}{5}+\frac{a_{3}k_{a}^{3}}%
{3}+a_{5}k_{a}-...)+\frac{1}{C^{3}}\left[  \frac{1}{\pi}\int\limits_{0}%
^{E_{a}}\delta(E^{\prime})\left(  E^{\prime}\right)  ^{2}dE^{\prime}-\frac
{1}{3}\sum_{n}\left(  E_{n}\right)  ^{3}\right]  ,\nonumber
\end{align}
where $E_{a}=Ck_{a}^{2}$ is an arbitrary energy value in the range where the
asymptotic approximation Eq. (20) can be used.

Eq. (36) may look nice but it is a bit inconvenient in practice.
Fortunately, a straightforward approach exists, which enables to ascertain the
coefficients in Eq. (35)\ more accurately and much more easily without any
direct reference to the phase shift. The approach in question is based on the
following integral representation for the Jost function \cite{Chadan}:%
\begin{equation}
F(k)=1+\frac{1}{C}\int\limits_{0}^{\infty}e^{ikr}V(r)\varphi(k,r)dr,
\end{equation}
where the regular solution $\varphi(k,r)$ can be calculated with the help of a
well-known iteration method. Namely, taking $\varphi^{(0)}(k,r)=\dfrac{\sin
kr}{k}$, and%
\begin{equation}
\varphi^{(n)}(k,r)=\frac{1}{C}\int\limits_{0}^{r}\dfrac{\sin k(r-r^{\prime}%
)}{k}V(r^{\prime})\varphi^{(n-1)}(k,r^{\prime})dr^{\prime},\text{ }n=1,2,...,
\end{equation}
the desired solution reads%
\begin{equation}
\varphi(k,r)=\sum_{n=0}^{\infty}\varphi^{(n)}(k,r).
\end{equation}

In Eq. (38) one can use integration by parts to calculate step-by-step the
terms $\varphi^{(1)}(k,r),$ $\varphi^{(2)}(k,r),$ $\varphi^{(3)}(k,r),...$,
and their contributions to the Jost function. Let us see, how to ascertain the
correct asymptotic formula for the function $g(k)=\dfrac{1}{\left\vert
F(k)\right\vert ^{2}}-1$ as $k\rightarrow\infty,$ which would include the
terms until $\sim1/k^{4}.$ Within this approximation%
\begin{align}
\varphi^{(1)}(k,r) &  =-\frac{\cos kr}{2Ck^{2}}W(r)+\frac{\sin kr}{4Ck^{3}%
}\left[  V(0)+V(r)\right]  +\frac{\cos kr}{8Ck^{4}}\left[  V^{\prime
}(r)-V^{\prime}(0)\right]  ,\text{ }W(r)\equiv\int\limits_{0}^{r}V(r^{\prime
})dr^{\prime},\\
\varphi^{(2)}(k,r) &  =-\frac{\sin kr\cdot\left[  W(r)\right]  ^{2}}%
{8C^{2}k^{2}}-\frac{\cos kr}{8C^{2}k^{4}}\left[
V(0)W(r)+V(r)W(r)+U(r)\right]  ,\text{ }U(r)\equiv\int\limits_{0}^{r}\left[
V(r^{\prime})\right]  ^{2}dr^{\prime},\nonumber\\
\varphi^{(3)}(k,r) &  =\frac{\cos kr\cdot\left[  W(r)\right]  ^{3}}%
{48C^{3}k^{4}}\nonumber
\end{align}
(all higher order terms can be ignored). Thereafter, using integration by
parts in Eq. (37), one comes to the following formulas for the real and
imaginary parts of the Jost function:%
\begin{align}
\operatorname{Re}F(k) &  =1+\frac{V(0)}{4Ck^{2}}-\frac{W^{2}}{8C^{2}k^{2}%
}-\frac{V^{\prime\prime}(0)}{16Ck^{4}}+\frac{1}{32C^{2}k^{4}}\left\{  5\left[
V(0)\right]  ^{2}-2V^{\prime}(0)W\right\}  -\\
&  -\frac{1}{32C^{3}k^{4}}\left[  V(0)W^{2}+2UW\right]  +\frac{W^{4}}%
{384C^{4}k^{4}},\nonumber
\end{align}%
\begin{equation}
\operatorname{Im}F(k)=\frac{W}{2Ck}+\frac{V^{\prime}(0)}{8Ck^{3}}+\frac
{1}{8C^{2}k^{3}}\left[  V(0)W+U\right]  -\frac{W^{3}}{48C^{3}k^{3}},
\end{equation}
where $W\equiv\int\limits_{0}^{\infty}V(r^{\prime})dr^{\prime}$ and
$U\equiv\int\limits_{0}^{\infty}\left[  V(r^{\prime})\right]  ^{2}dr^{\prime
}.$ Quite surprisingly, when calculating $\left\vert F(k)\right\vert
^{2}=\left[  \operatorname{Re}F(k)\right]  ^{2}+\left[  \operatorname{Im}%
F(k)\right]  ^{2}$, all troublesome terms will cancel out, resulting in nice
expressions for both quantities of interest:%
\begin{equation}
\left\vert F(k)\right\vert ^{2}=1+\frac{V(0)}{2Ck^{2}}+\frac{3\left[
V(0)\right]  ^{2}}{8C^{2}k^{4}}-\frac{V^{\prime\prime}(0)}{8Ck^{4}},
\end{equation}%
\begin{equation}
g(E)=\frac{V(0)}{2E}+\frac{\left[  V(0)\right]  ^{2}-CV^{\prime\prime}%
(0)}{8E^{2}}.
\end{equation}
Comparing Eqs. (35) and (44), one finds that the coefficients%
\begin{equation}
a_{2}=\frac{V(0)}{4C},\text{ }a_{4}=\frac{2\left[  V(0)\right]  ^{2}%
-CV^{\prime\prime}(0)}{16C^{2}}%
\end{equation}
do not depend on the phase shift, but are directly related to the potential
and its second derivative at the zero point, both these quantities being
finite according to the starting idea of the approach. Thus we have described
all properties of the Jost function in the frame of the proposed approach, and
are now prepared to start calculation of the potential itself.

\section{Conclusion}

In this paper we proposed as if an "inversed" approach to the quantum
mechanical inverse problem. Starting from a known reference potential, one can
calculate the important spectral characteristics of the system, the phase
shift and the Jost function, which are almost unattainable in a real
experiment. On one hand, the reference potential has to be realistic enough to
be used as a zeroth approximation to the real potential. Only in this case
there is a chance to construct a Bargmann potential that would be even more
realistic, for example, whose discrete energy levels would exactly fit with
the actually observed ones. On the other hand, we suggest to choose a
reference potential that would be exactly solvable, in the sense that its
energy eigenvalue problem can be solved to any desired accuracy with the help
of solely analytic means. To this end, as demonstrated in Sections 2 and 3, a
multi-component potential composed of smoothly joined Morse-type pieces is
especially suitable. In particular, we would like to highlight the usefulness
of the pseudo-Morse approximation for the small distances (and high energies)
region of the potential. In this paper we only used a single pseudo-Morse
component, which stretches until the zero point $r=0$. As demonstrated
elsewhere \cite{Selg1, Selg2}, adding more pseudo-Morse components does not
bring along any serious problems, so one can include just as many components
of this type as he considers reasonable.

The special analytic form of the reference potential enables to ascertain the
phase shift analytically up to the energies $E\lessapprox V(0)$ (note that an
arbitrarily large value for $V(0)$ can be taken, if one introduces a
sufficient number of pseudo-Morse components). As the asymptotic behavior of
the phase shift can be determined from general physical considerations, it is
possible to directly demonstrate full agreement with the famous Levinson
theorem \cite{Levinson1}, as well as to ascertain the full energy dependence
of the Jost function. Provided with this important input information, one can
attack the real computational-technical problem, trying to solve the main
integral equation in the frame of the Gelfand-Levitan \cite{GL}, Marchenko 
\cite{Marchenko2} or Krein \cite{Krein2} method. This will be the subject of
the next paper in this series \cite{SelgII}.

\section*{Acknowledgement}

The research described in this paper has been supported by Grants No 5863 and
4508 from the Estonian Science Foundation.

\pagebreak

\section*{Figure captions}

\begin{enumerate}

\item[Fig. 1.] Three-component model potential for the system (Xe$_{2}$ in
ground electronic state) investigated. Note that the same potential curve is
depicted in both graphs (the lower one starts where the upper one ends), but
essentially different energy scales are used for them. All components have the
well-known analytic form of the Morse potential, but the ordinary Morse
approximation is used only in the central range $r\in\left[  X_{1}%
,X_{2}\right]  $ (see the explanations in Section 2). The parameters of the
components as well as the calculated discrete energy levels (24 in total) are
given in Table 1.

\item[Fig. 2.] Direct demonstration of the Levinson theorem ($\delta
(0)-\delta(\infty)=n\pi$) for the model system studied. As needed,
$\delta(0)=24\pi$, since the system has 24 bound states. At $E=$ 3.146294 meV,
the phase shift passes a zero, and then remains negative, very slowly
approaching the limit ($\delta(\infty)=0$) as $E\rightarrow\infty.$ The left
side-inset shows the nearly linear energy dependence as $E\rightarrow0$, in
full agreement with Eq. (22). In the right-side inset one can see that the
phase curve has an inflection point near $E=V(0)$.

\item[Fig. 3.] A demonstration of the exellent agreement between the
numerically calculated phase shifts and the general asymptotic formula, Eq.
(20), as $E\rightarrow\infty$.

\item[Fig. 4.] Demonstration of the results of calculating the Jost function
(in fact, its modulus) for the reference potential. Again, as in the case of
Fig. 1, different parts of the same curve are shown in the graphs and their
insets, but rather different scales are used (note that in some cases the
scale is logarithmic). The information gathered into the upper graph would be
useful, if one's aim is to construct another potential whose Jost function
only slightly differs from the initial one. The lower graph demonstrates how
abruptly the curve turns from practically vertical to nearly horizontal in the
vicinity of the "critical" energy value $E=V(0)$, while this dramatic change
is absolutely invisible in the scale used for the ordinate axis in the upper
graph. The asymptotic formula Eq. (35) is used to calculate the Jost function
for $k\geq k_{a}=75000$ \AA \ (which corresponds to about 1.8*10$^{5}$ eV)
with coefficients given by Eq. (45).

\item[Fig. 5.] Visualization of the characteristic function $g(k)$ (given gy
Eq. (33)), which determines the kernel of the Gelfand-Levitan integral
equation. As can be seen, $g(k)=1$ (with very high precision) if $k\lessapprox
k_{0}=\sqrt{V(0)/C}$ (see Fig. 4 to undestand the reason for this), then
starts to decrease, gradually "breaking free" from the potential field, and
$g(k)\rightarrow0$ as $k\rightarrow\infty$. In spite of the seeming simplicity
of the energy dependence, the whole curve has to be (and has been) calculated
very accurately up to very high energies, in order to accurately ascertain the
potential. The high-energy part of the curve ($k\geq k_{a}$) has been
calculated according to Eq. (44).
\end{enumerate}

\pagebreak

\begin{table}[ptb]
\caption{Parameters of the three-component reference potential for Xe$_{2}$
(first column) and its discrete energy eigenvalues $E_{n}$.}%
\label{table1}%
\begin{tabular}
[c]{|l|l|l|}\hline\hline
Units used: & $n$ & $\ \ E_{n}$ (meV)\\\cline{2-3}%
$V_{k}$ ($k=0,1,2$) - meV & 0 & -23.043278\\
$D_{k}$ - meV & 1 & -20.618294\\
$\alpha_{k}$ - 1/\AA  & 2 & -18.347033\\
$r_{k},$ $X_{1}$, $X_{2}$ - \AA  & 3 & -16.229494\\
& 4 & -14.266840\\
$V_{0}=$ -38.865765625809 & 5 & -12.457771\\
$D_{0}=$ 0.02078293632204 & 6 & -10.802778\\
$R_{0}=$ 6.23549082364147 & 7 & \ -9.301735\\
$\alpha_{0}=$ 1.61583465987087 & 8 & \ -7.954433\\
& 9 & \ -6.760559\\
$V_{1}=$ -24.3134155 & 10 & \ -5.718592\\
$D_{1}=$ 21.629 & 11 & \ -4.817731\\
$R_{1}=$ 1.5537 & 12 & \ -4.028060\\
$\alpha_{1}=$ 4.3634 & 13 & \ -3.328990\\
& 14 & \ -2.712825\\
$V_{2}=$ 0.0155845 & 15 & \ -2.171217\\
$D_{2}=$ $-V_{2}$ & 16 & \ -1.698210\\
$R_{2}=$ 0.3755186 & 17 & \ -1.289229\\
$\alpha_{2}=$ 14.05149 & 18 & \ -0.940193\\
& 19 & \ -0.647734\\
$X_{1}=$ 4.00000 & 20 & \ -0.409167\\
$X_{2}=$ 6.05149 & 21 & \ -0.222466\\
& 22 & \ -0.086525\\
& 23 & \ -0.002574\\\hline\hline
\end{tabular}
\end{table}

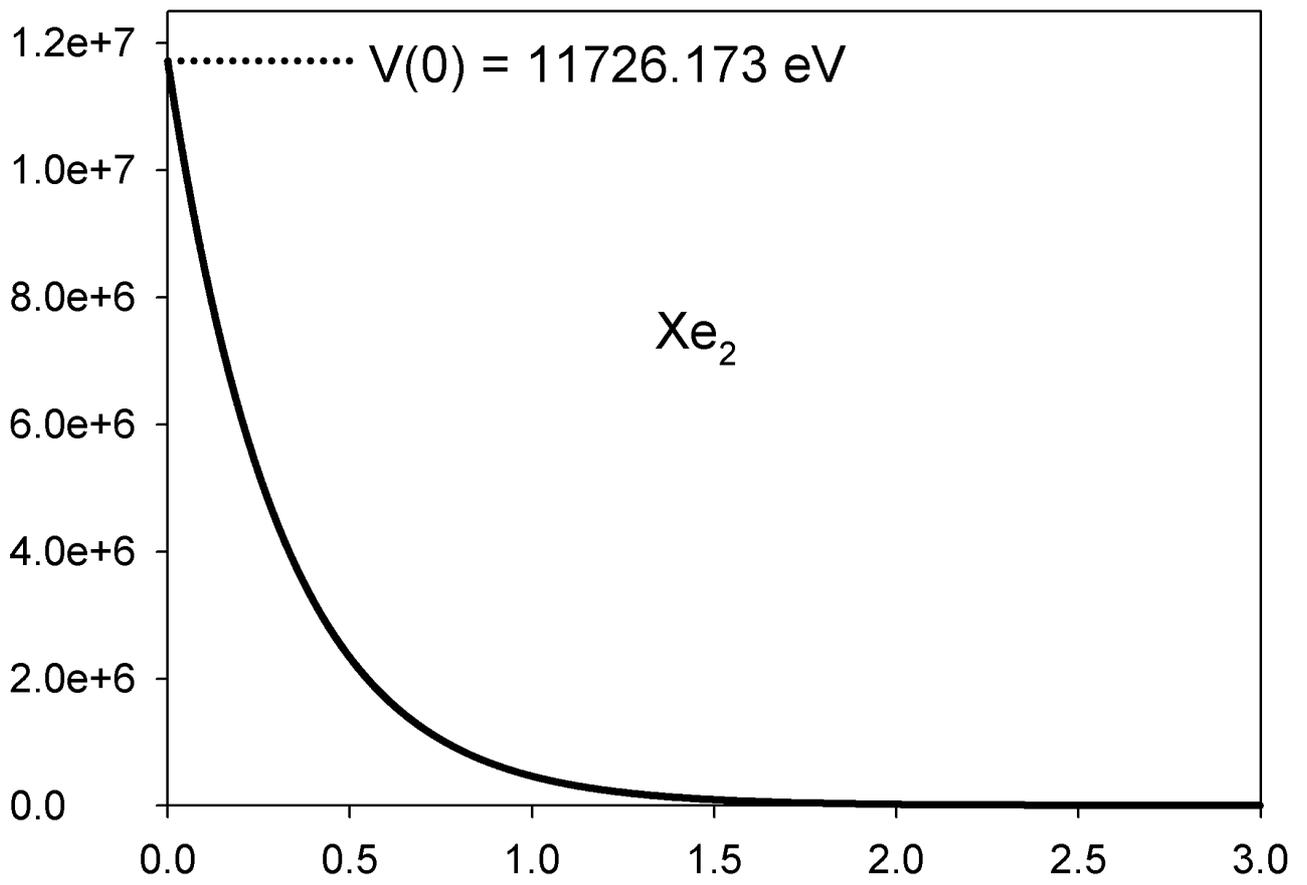
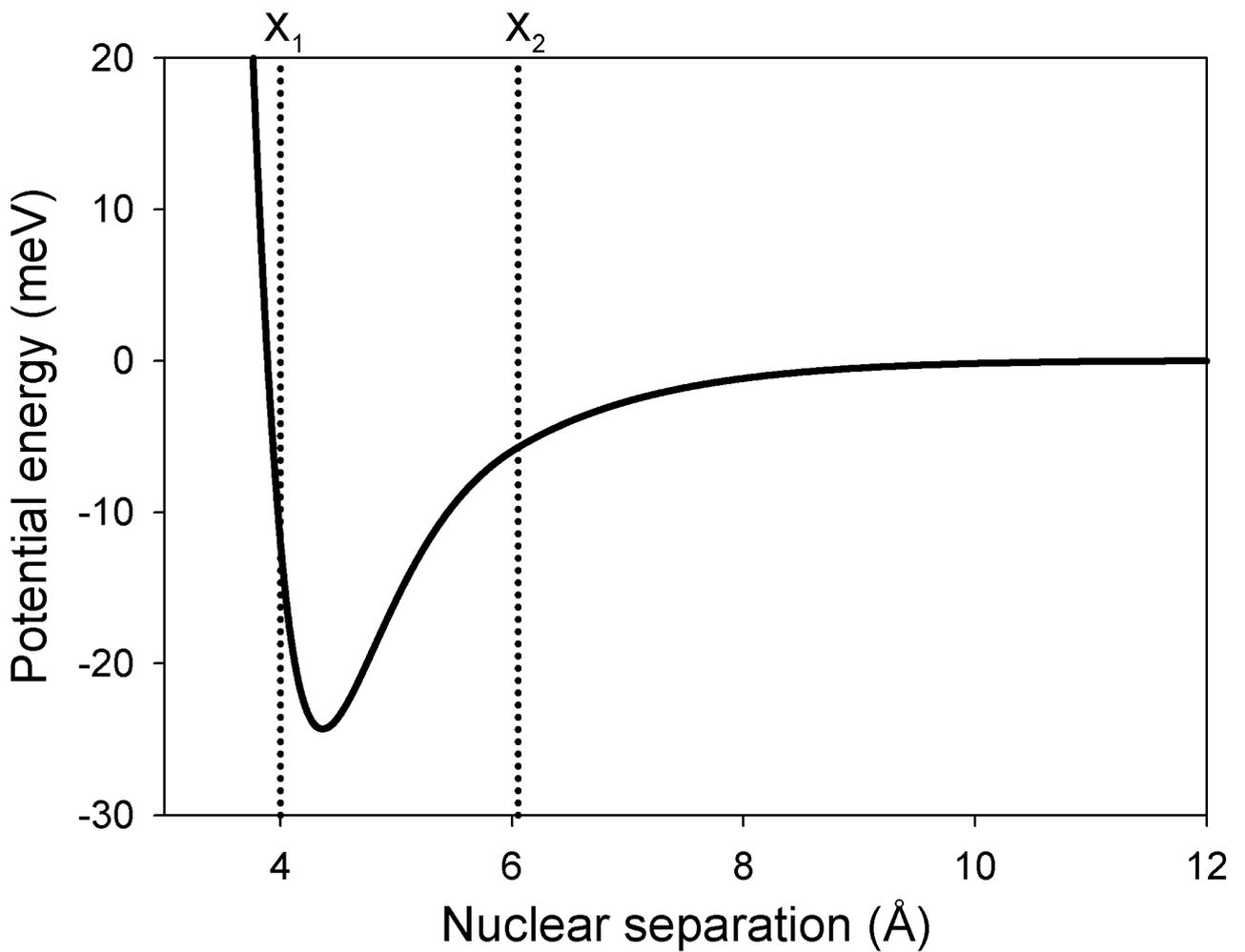
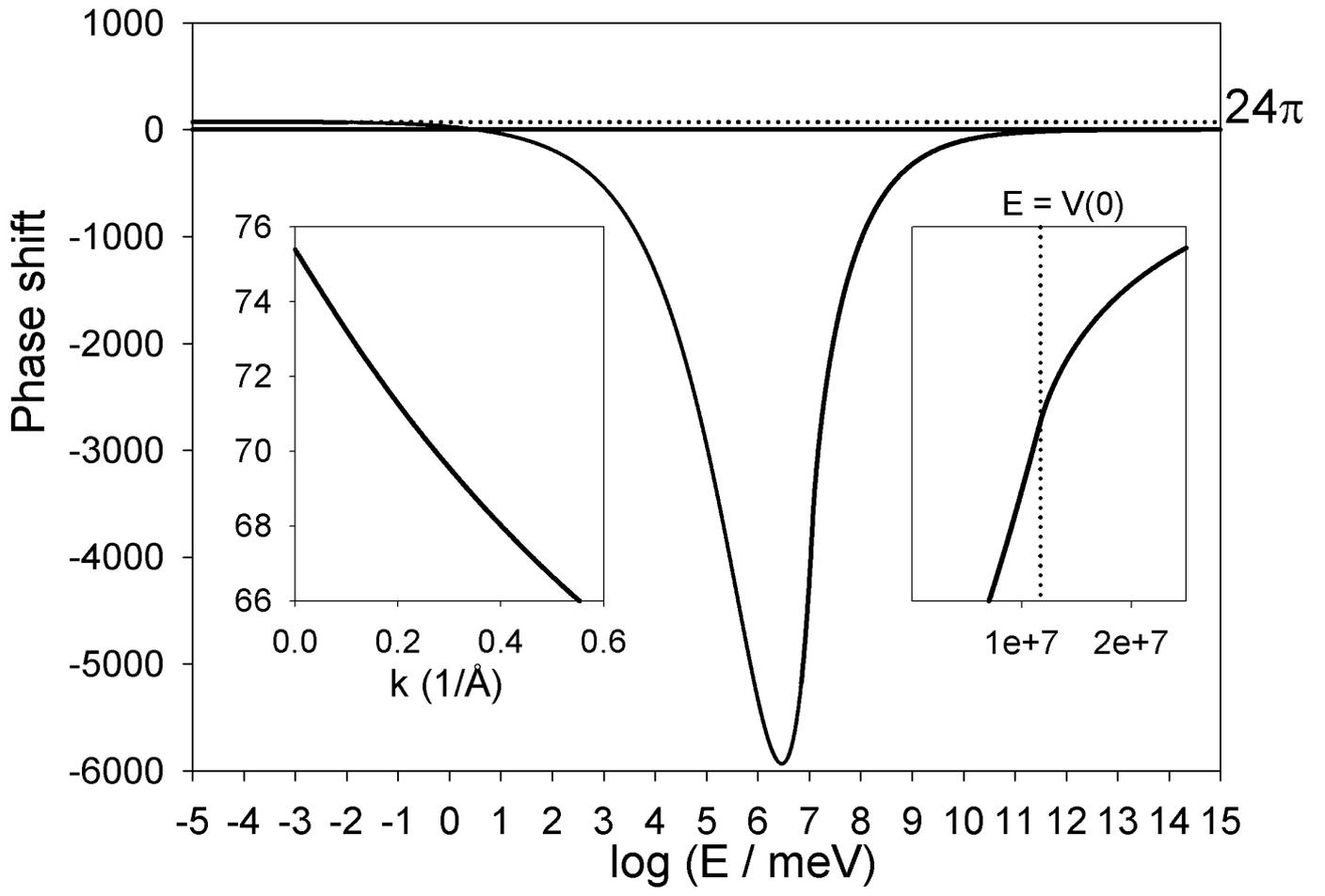
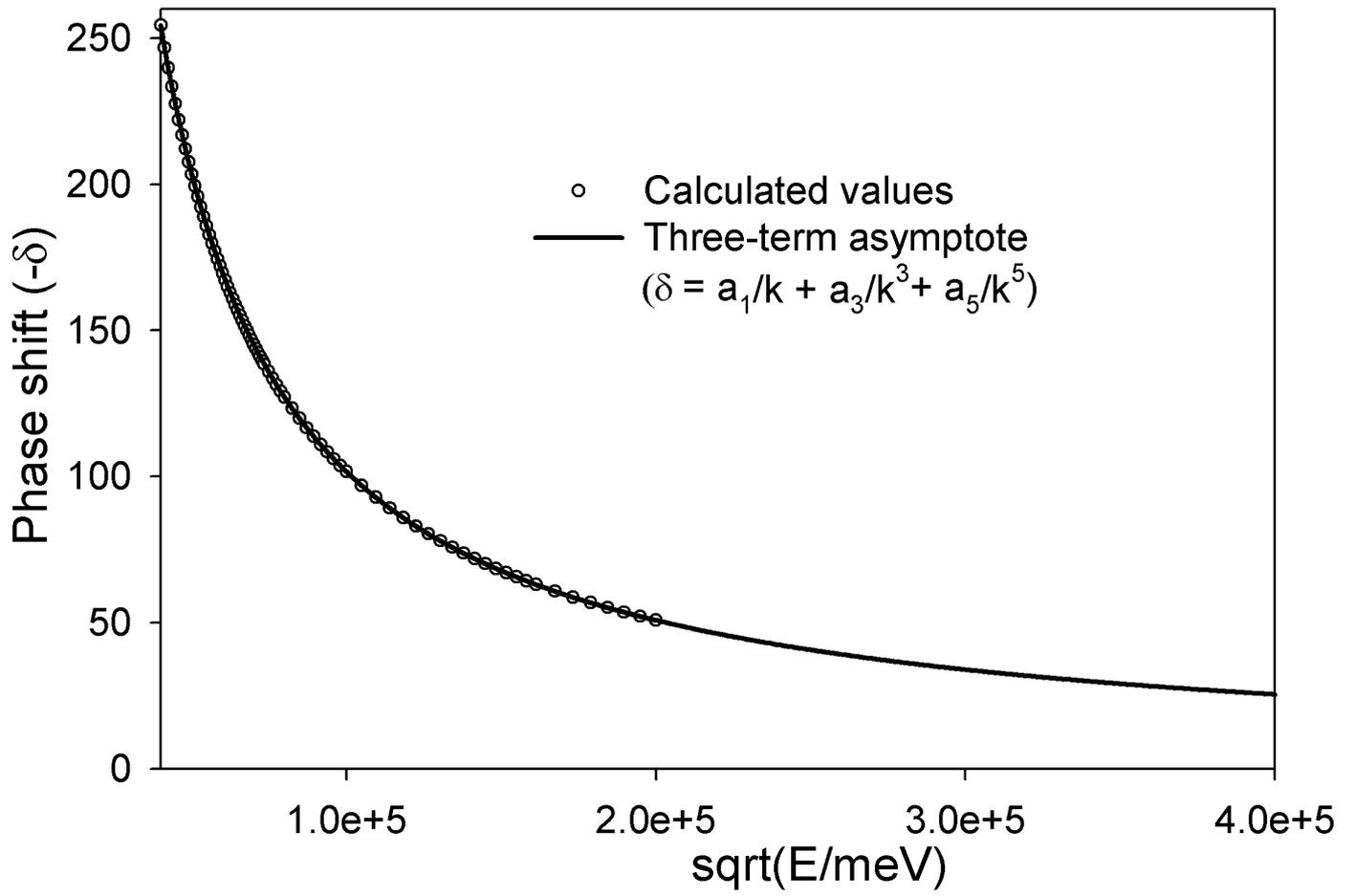
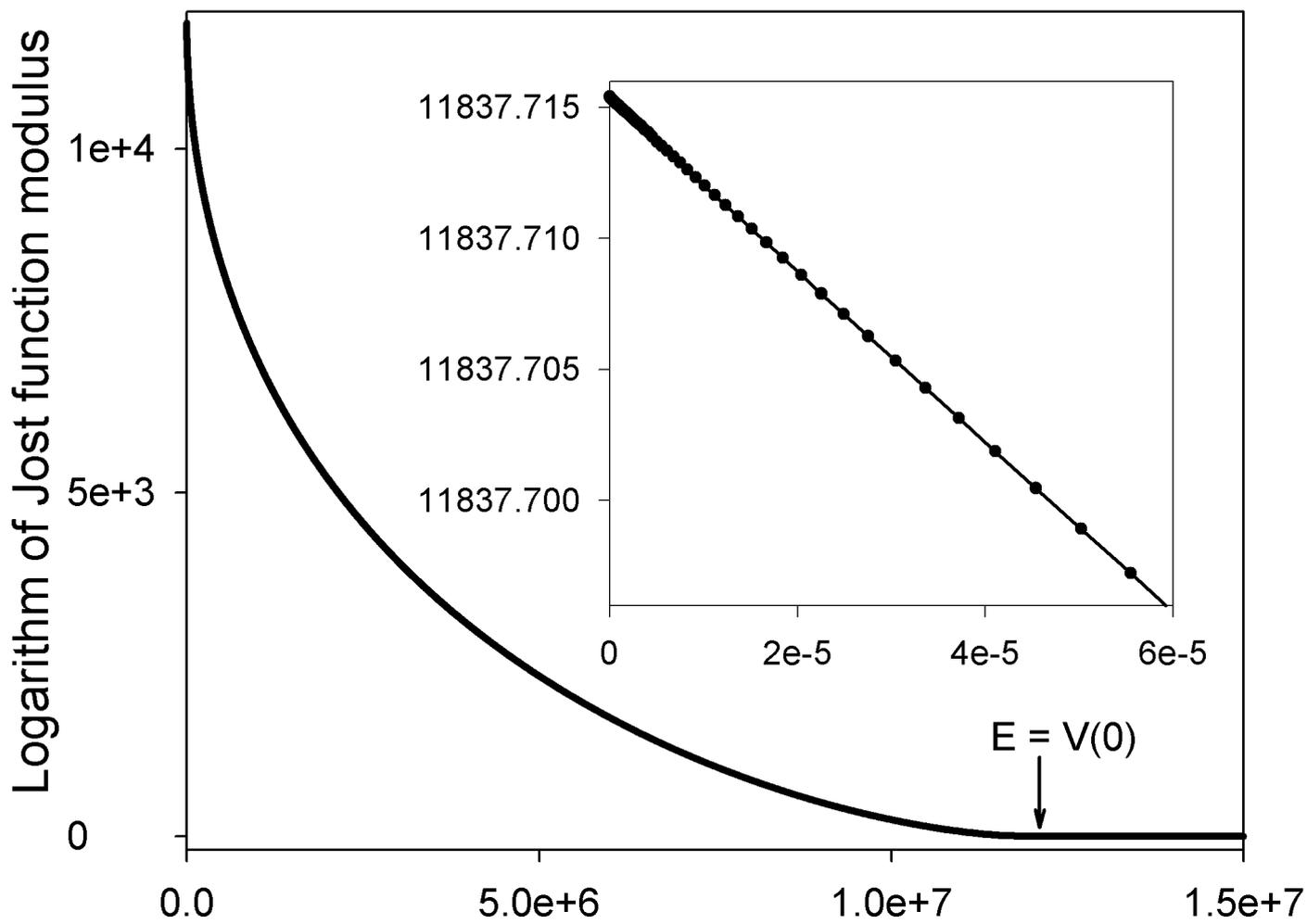
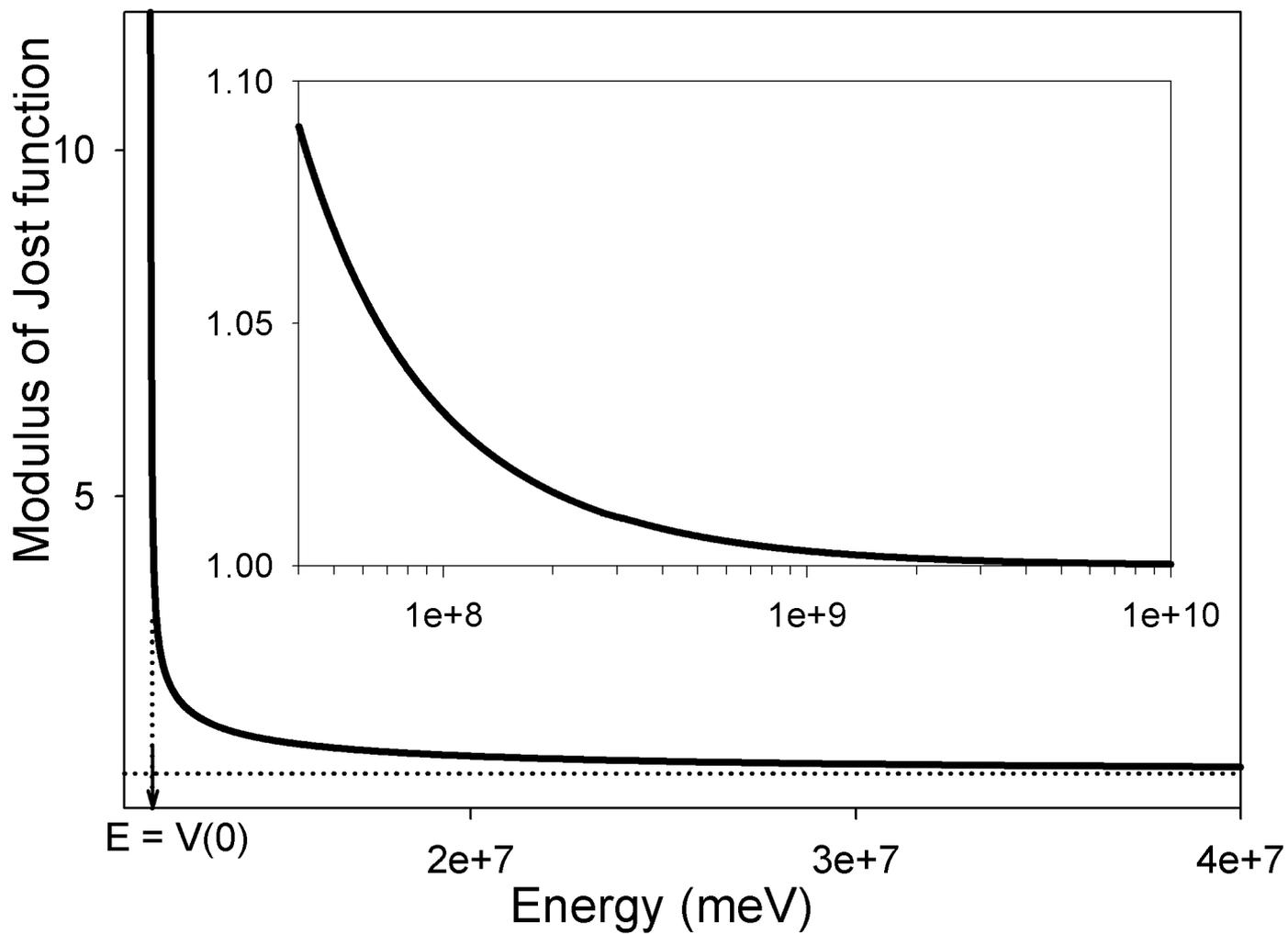
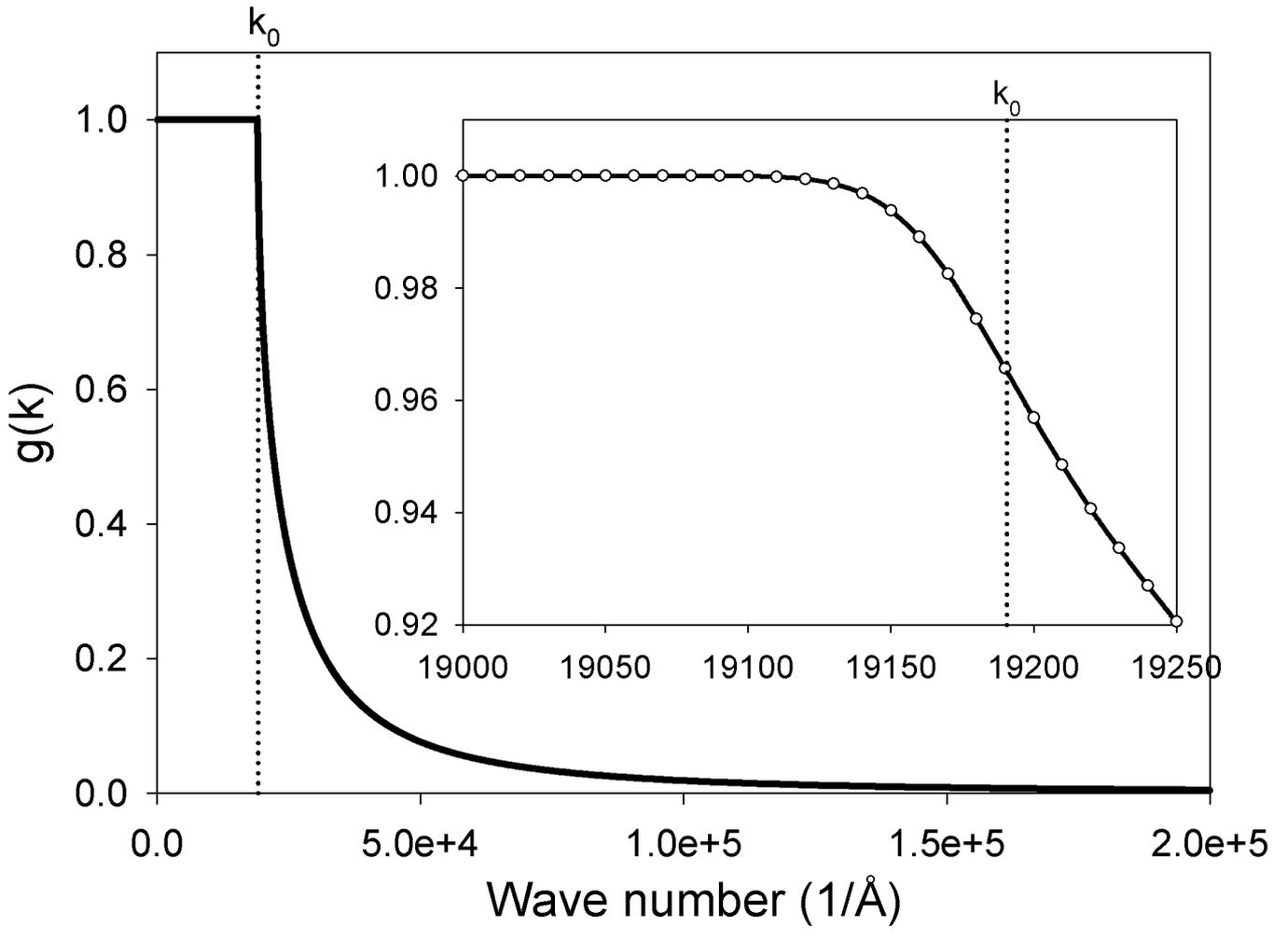
\end{document}